\documentclass[prr,twocolumn,superscriptaddress,nofootinbib,longbibliography]{revtex4-2}
\usepackage{dcolumn}
\usepackage{bm}
\usepackage{graphicx}
\usepackage{amsfonts}
\usepackage{bbm}
\usepackage{dsfont}
\usepackage{float}
\usepackage{graphicx}
\usepackage{bbm}
\usepackage{amssymb}
\usepackage{multibib}
\usepackage{color}
\usepackage{enumerate}
\usepackage{amsmath}
\usepackage{amstext}
\usepackage{latexsym}
\usepackage{braket}
\usepackage{appendix}
\usepackage{textcomp}
\usepackage[usenames,dvipsnames]{xcolor}
\usepackage[colorlinks=true,citecolor=Blue,linkcolor=RubineRed,urlcolor=Blue]{hyperref}
\usepackage{mathtools}
\usepackage{lipsum}
\usepackage{grffile}
\usepackage{subfigure}
\usepackage{overpic}

\begin{document}

\title{Decoherence rate in random Lindblad dynamics}
\date{\today}

\author{Yifeng Yang\href{https://orcid.org/0009-0009-2446-6378}
{\includegraphics[scale=0.05]{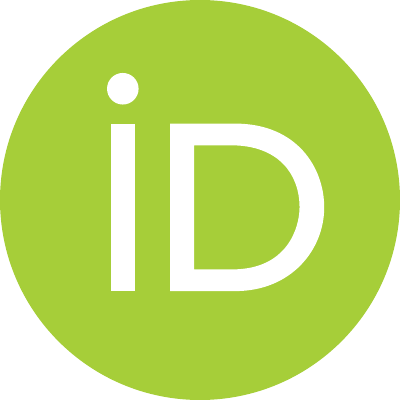}}}
\affiliation{School of Physical Science and Technology, Soochow University, Suzhou 215006, China}
\author{Zhenyu Xu\href{https://orcid.org/0000-0003-1049-6700}
{\includegraphics[scale=0.05]{orcidid.pdf}}}
\affiliation{School of Physical Science and Technology, Soochow University, Suzhou 215006, China}
\author{Adolfo del Campo\href{https://orcid.org/0000-0003-2219-2851}{\includegraphics[scale=0.05]{orcidid.pdf}}}
\affiliation{Department  of  Physics  and  Materials  Science,  University  of  Luxembourg,  L-1511  Luxembourg, G. D.  Luxembourg}
\affiliation{Donostia International Physics Center,  E-20018 San Sebasti\'an, Spain}

\begin{abstract}

Open quantum systems undergo decoherence, which is responsible for the transition from quantum to classical behavior. 
The time scale in which decoherence takes place can be analyzed using upper limits to its rate. We examine the dynamics of open chaotic quantum systems governed by random Lindblad operators sourced from Gaussian and Ginibre ensembles with Wigner-Dyson symmetry classes. In these systems, the ensemble-averaged purity decays monotonically as a function of time. This decay is governed by the decoherence rate, which is upper-bounded by the dimension of their Hilbert space and is independent of the ensemble symmetry. These findings hold upon mixing different ensembles, indicating the universal character of the decoherence rate limit. 
Moreover, our findings reveal that open chaotic quantum systems governed by random Lindbladians tend to exhibit the most rapid decoherence, regardless of the initial state. This phenomenon is associated with the concentration of the decoherence rate near its upper bound. 
Our work identifies primary features of decoherence in dissipative quantum chaos, with applications ranging from quantum foundations to high-energy physics and quantum technologies.

\end{abstract}

\maketitle

\section{Introduction}
Any quantum system is embedded in a surrounding environment.  The system-environment interaction results in the build-up of quantum correlations and leads to quantum decoherence, a process critical for comprehending the quantum-to-classical transition \cite{Zurek81,Zurek82,Zurek2003,Schlosshauer2019}. 
While decoherence is generally viewed as a fast process, the rate at which it occurs remains a topic of debate \cite{schlosshauer2007}. A thorough understanding of decoherence rates is relevant to quantum foundations as well as quantum technologies, including quantum metrology, control, simulation, and computation, among others. 
Early theoretical studies focused on quantum Brownian motion \cite{Zurek91,Zurek93}. 
In the high-temperature limit, when the dynamics is governed by dephasing in real space, an elegant formula due to Zurek estimates the decoherence rate $D$ of a wavepacket in terms of its dispersion $\Delta x$ in the coordinate representation in units of the de Broglie wavelength $\lambda_{\rm dB}$, i.e., $D\propto (\Delta X/\lambda_{\rm dB})^2$ \cite{Zurek91,ZurekPaz94}. This expression follows from the explicit time dependence of the off-diagonal coherences of the density matrix describing the quantum state of the open system \cite{Zurek91}. 

An important step forward relied on a rigorous basis-independent definition of the decoherence rate for Markovian open quantum systems, in which the quantum state fulfills a master equation of the Gorini-Kossakowski-Sudarshan-Lindblad (GKSL) form \cite{GKS,Lindblad}. The short-time behavior of the purity, the linear entropy, and the fidelity identify the decoherence rate in terms of the covariance of the Lindblad operators
\cite{Zurek93,Lidar98,Shimizu02,Chenu17,Beau17,Xu19,Xu2021,GarciaPintos21,Korbicz21}. Zurek's formula for quantum Brownian motion is obtained as a special case \cite{Zurek93,Beau17}. The availability of a universal expression for Markovian open systems has made possible the study of scenarios characterized by extreme decoherence \cite{Xu19} and the exploration of the fundamental limits to the pace of decoherence, leading to the introduction of decoherence rate limits \cite{Chenu17,Beau17,Xu19}.

Exploring open chaotic quantum systems offers novel insights into decoherence \cite{Habib1998,Karkuszewski02,Xu19}. According to the Bohigas-Giannoni-Schmit conjecture, any Hamiltonian quantum system that displays chaotic behavior in its classical limit is expected to have spectral statistics described by random matrix theory \cite{BGS1984}. 
In physics, random matrix theory was introduced by Wigner to characterize the spectral density of heavy nuclear atoms \cite{Wigner51,Wigner55}.  Ensuing work by Dyson developed a symmetry classification distinguishing complex Hermitian, or real symmetric, and quaternion self-dual random matrices  \cite{Dyson62a,Dyson62b,Dyson62c,Dyson1962}. This classification, known as the threefold way, led to the identification of orthogonal (O), unitary (U),  and symplectic (S) ensembles. When they are associated with a Hamiltonian probability density function of the form $P(H)\propto\exp(-{\rm tr} H^2/2)$ they are known as Gaussian ensembles and labeled as GXE (${\rm X=O,U,S}$) according to the threefold way. More elaborated classification schemes have been put forward \cite{AltlandZirnbauer97}. 
To date, random matrix theory has found extensive applications across various domains of physics, and notably,  in chaotic quantum systems \cite{MethaBook,Haake,BookVivo2017}.  Random matrix Hamiltonians constitute a paradigmatic framework for the description of complex isolated systems.

Embracing  a similar approach for the description of complex open quantum systems  has led to the study of  
 random Lindbladians \cite{Xu19,Can19,Denisov19,Can2019PRL,Sa_2020,Wang2020,Lucas2020PRX,delCampo2020,Xu2021,HuiZhai2021,Cornelius2022,Roubeas2023,Lucas2023PRX,Kawabata2023,Xiao2023PRXQ,Apollonas2023PRA,zhou2023,ferrari2023arXiv}.  
The Lindbladian is the generator of evolution in a Markovian quantum system \cite{Lindblad}. Ensembles of random Lindbladians fall into different classes distinguished by their symmetries \cite{Lucas2023PRX,Kawabata2023,GarciaGarcia23}. We focused on the decoherence dynamics associated with random Lindbladians.
In this context, fluctuating chaotic quantum systems, characterized by random Lindblad operators described by random matrices sampled from GUE, exhibit the fastest decoherence rates reported to date \cite{Xu19}. This rate of decoherence scales exponentially with the number of particles, i.e., it is linear in the dimension of the Hilbert space. This is in stark contrast with the dependence found in systems with fluctuating $k$-body interactions, that is polynomial in the system size \cite{Chenu17,Beau17metrology}.

The use of GUE ensembles in random Lindbladians is associated with Hermitian Lindblad operators, without time-reversal symmetry, when the dissipative dynamics reduces to dephasing. 
What decoherence rates govern open systems with  Lindblad operators with other symmetries,  as in the GOE and GSE, which are distinguished by the presence and absence of spin-rotation invariance? What is the characteristic decoherence dynamics under Lindbladians sampled from the Ginibre ensembles of non-Hermitian matrices \cite{ginibre1965statistical}? Is it possible for the decoherence rates to surpass those known in GUE? What is the dependence on the initial state?

We address the posed questions by examining random Lindblad operators using Hermitian matrices from GXEs and non-Hermitian matrices from Ginibre ensembles (GinXEs) (${\rm X=O,U,S}$). Employing Haar averaging in conjunction with the spectral density function and the two-point correlation spectral function, we analytically obtain the decoherence rate averaged over these ensembles. This decoherence rate governs the ensemble-averaged purity decay as a function of time. Our analytical and numerical results demonstrate that the decoherence rate across all the examined scenarios is upper bounded by the dimension of the system's Hilbert space, establishing a universal limit.

Our study further reveals that when the initial state is uniformly sampled—from a pure state to a maximally mixed state—across a one-parameter family, the decoherence rate predominantly clusters around the maximal rate. We term this phenomenon \textit{concentration of the decoherence rate}. This suggests that quantum chaotic open systems predominantly follow the fastest decoherence pathway, irrespective of the initial state configuration.

The structure of the remainder of this paper is organized as follows: In Sec. \ref{DRdef}, we introduce the definition of the decoherence rate in the context of the short-time behavior of purity. Sections \ref{SecGXE} and \ref{SecGinXE} (and Appendixes \ref{app-0}--\ref{app-B}) detail the analytical calculation of the decoherence rate, averaged over GXEs and GinXEs, complemented by numerical simulations. 
%Then, an intriguing phenomenon concerning t
The concentration of the decoherence rate is analyzed in Sec. \ref{SecCP}. The symmetry crossover between the GXE and GYE holds significance in the context of symmetry breaking in quantum chromodynamics \cite{Kanazawa2020,KANAZAWAPLB}. We briefly discuss the decoherence rate over this symmetry crossover and conclude in Sec. \ref{SecDC}.

\section{The decoherence rate and the random Lindblad operator}\label{DRdef}	
The master equation encodes the time evolution of the quantum state of an open system as a first-order differential equation in which the Lindbladian $\mathcal{L}$ acts as the generator of the evolution, i.e., $\dot{\rho}_t=\mathcal{L}\left(\rho_t\right)$. In the Markovian regime, the generator of the dynamical semigroup admits the Gorini-Kossakowski-Sudarshan-Lindblad (GKSL) form \cite{GKS,Lindblad}
\begin{eqnarray}
  \dot{\rho}_t&=&\mathcal{L}\left(\rho_t\right)=-\frac{i}{\hbar}[H, \rho_t]+\mathcal{L}_D(\rho_t),\\
  \mathcal{L}_D(\rho_t)&=&\sum_{\alpha}\gamma_{\alpha}\left( L_{\alpha}\rho_t L_{\alpha}^{\dagger}-\frac{1}{2}\left\{ L_{\alpha}^{\dagger}L_{\alpha},\rho_t\right\}\right).
\end{eqnarray}      
%with $\mathcal{L}_D(\rho_t)=\sum_{\alpha}\gamma_{\alpha}\left( L_{\alpha}\rho_t L_{\alpha}^{\dagger}-\frac{1}{2}\left\{ L_{\alpha}^{\dagger}L_{\alpha},\rho_t\right\}\right)$. 
Here, the dissipator $\mathcal{L}_D$ is composed of the Lindblad operators $L_{\alpha}$ and the damping rates $\gamma_{\alpha}\geq0$, and it involves the anticommutator $\{ \cdot \}$. The set of Lindblad operators characterizes the system's dissipative dynamics, encapsulating the environmental influence on the system. The formulation of these Lindblad operators can be either Hermitian or non-Hermitian, depending on the type of decoherence \cite{BreuerBook,BookRivasHuelga}.

The purity, denoted as $P_{t}=\operatorname{tr}(\rho_{t}^{2})$, can be used to quantify decoherence and diminishes as the quantum state $\rho_{t}$ is increasingly mixed \cite{nielsen_chuang_2010,jozsa1994fidelity}. When $\rho_{t}$ is a pure state, $P_{t}=1$. Conversely, in a maximally mixed state, the purity is the inverse of the Hilbert space dimension $N$. The timescale governing the non-unitary nature of the evolution can be inferred from the short-time behavior of the purity, given by $P_{t}/P_0=1-D \cdot t+\mathrm{O}\left(t^2\right)$ \cite{Lidar98,Shimizu02,Beau17,Xu19,Xu2021}, where 
\begin{align}
		D=\frac{2\sum_{\alpha}\gamma_{\alpha}\left[ \operatorname{tr}(\rho_{0}^2 L_{\alpha}^{\dagger}L_{\alpha})-\operatorname{tr}(\rho_{0}L_{\alpha}^{\dagger}\rho_{0}L_{\alpha})\right]}{P_0}, \label{D}
\end{align}
is called the decoherence rate. It is exclusively governed by $\mathcal{L}_D$ and independent of the system Hamiltonian \cite{Beau17,Xu19}. As already indicated, modeling complex open quantum systems is efficient in terms of random Lindbladians when Lindblad operators are chosen as random matrices subject to given symmetries \cite{Xu19,Can19,Denisov19,Can2019PRL,Sa_2020,Wang2020,Lucas2020PRX,delCampo2020,Xu2021,Cornelius2022,Roubeas2023,Lucas2023PRX,Kawabata2023,Apollonas2023PRA,zhou2023}. In this paper, we consider $L_{\alpha}$ as a random matrix with elements drawn from Gaussian distributions, encompassing both Hermitian and non-Hermitian scenarios. In particular, we consider the Gaussian and Ginibre ensembles. Additionally, we focus on the most widely used Wigner-Dyson symmetry classes: the orthogonal, unitary, and symplectic symmetries \cite{MethaBook}. The decoherence rate, when averaged across the ensemble, results in (see details in Appendix \ref{app-0})
\begin{equation}
  \langle D\rangle=2 \Gamma \frac{N P_0-1}{\left(N^2-1\right) P_0}\left[\left\langle\operatorname{tr}\left(L^{\dagger} L\right)\right\rangle-\frac{1}{N}\left\langle\left(\operatorname{tr} L^{\dagger}\right)(\operatorname{tr} L)\right\rangle\right], \label{<D>}
\end{equation} 
where $\Gamma=\sum_\alpha \gamma_\alpha$. It is important to note that while the subscript $\alpha$ of $L_\alpha$ can be reinstated in Eq. (\ref{<D>}), this is not necessary, as the ensemble averages across different $L_\alpha$ variants are identical. Furthermore, for the sake of simplicity, the Lindblad operator $L$ will be directly sampled from GXE or GinXE. The choice of a traceless Lindblad operator $\tilde{L}$ is ensured by a shift $\tilde{L}=L-\frac{\mathbbm{1}}{N}\operatorname{tr}L$, thereby guaranteeing its equivalence to $L$ (see proof in Appendix \ref{app-C}). Obviously, Eq. (\ref{<D>}) is upper bounded by $\langle D\rangle \leq \langle D_L\rangle$, with
\begin{equation}
  \langle D_L \rangle =  \frac{2 \Gamma}{N+1}\left[\left\langle\operatorname{tr}\left(L^{\dagger} L\right)\right\rangle-\frac{1}{N}\left\langle\left(\operatorname{tr} L^{\dagger}\right)(\operatorname{tr} L)\right\rangle\right]. \label{upD}
\end{equation}	
The equality in $\langle D\rangle \leq \langle D_L \rangle$ is achieved when the initial state is pure. In subsequent sections, we will detail the computation of the upper bound of decoherence rate in Eq. (\ref{upD}) for both GXE and GinXE, respectively.

\section{The decoherence rate with GXE} \label{SecGXE}
Consider a Hilbert space of dimension $N$.  
We define a random matrix $(a_{mn})_{N \times N}$ as belonging to the Gaussian ensembles if it is a real symmetric matrix with diagonal entries $a_{mm}\in \mathcal{N}(0,\sigma^{2})$, where $\sigma$ represents the standard deviation. The off-diagonal entries, i.e., the elements with $m\neq n$, are given by $a_{mn}=e^{0}_{mn}$ for GOE, $b_{mn}=e^{0}_{mn}+ie^{1}_{mn}$ for GUE, and $e_{mn}=e^{0}_{mn}+ie^{1}_{mn}+je^{2}_{mn}+ke^{3}_{mn}$ for GSE, respectively. Here, $e^{l}_{mn}\in\mathcal{N}(0,\sigma^{2}/2)$ ($l=0,1,2,3$), and $i$, $j$, $k$ are the basis elements for a quaternion. In this section, we consider $L_\alpha$ directly sampled from GXE, thus $L^\dagger=L$.

For GUE, orthogonal polynomials are widely employed to evaluate the spectral density function and two-point correlation function, i.e., $\varrho_{\operatorname{GUE}}(x)=\sum_{l=0}^{N-1}\phi_{l}(x)^{2}$ and $\varrho_{\operatorname{GUE}}(x,y)=\varrho_{\operatorname{GUE}}(x)\varrho_{\operatorname{GUE}}(y)-\sum_{m,n=0}^{N-1}\phi_{m}(x)\phi_{n}(x)\phi_{m}(y)\phi_{n}(y)$, where $\phi_{l}(x)$ are orthogonal polynomials \cite{MethaBook,Xu19}. However, when considering GOE, the orthogonal polynomials transform into skew-orthogonal polynomials, posing challenges for exact calculations at finite $N$. We employ an approximate method using formulas suitable for large dimensions for simplicity. For example, in the limit of large dimensions, the spectral density function of the Gaussian ensemble follows Wigner's semicircle distribution \cite{MethaBook}
\begin{equation}
		\varrho_{\operatorname{GOE}}(x)=\frac{\sqrt{N}}{\sigma\pi}\sqrt{1-\left(\frac{x}{2\sigma\sqrt{N}}\right)^{2}}. \label{pdf}
\end{equation}	
Consequently, the first term on the right-hand side of Eq. (\ref{upD}) is expressed as
\begin{align}
		\langle \operatorname{tr}(L^{2})\rangle_{\operatorname{GOE}}&=\int_{-2\sigma\sqrt{N}}^{2\sigma\sqrt{N}} x^{2}\varrho_{\operatorname{GOE}}(x)dx=\sigma^{2}N^{2}.
\end{align}
To calculate the second term on the right-hand side of Eq. (\ref{upD}), we divide it into two parts	
\begin{equation}
			\langle\operatorname{tr}(L)^{2}\rangle_{\operatorname{GOE}}=\int x^{2}\varrho_{\operatorname{GOE}}(x)dx+\int xy\varrho_{\operatorname{GOE}}(x,y)dxdy,
\end{equation}
where $\varrho_{\operatorname{GOE}}(x,y)=\varrho_{\operatorname{GOE}}(x)\varrho_{\operatorname{GOE}}(y)+\varrho^{c}_{\operatorname{GOE}}(x,y)$. Because of the symmetry of the spectral density function, $\int x\varrho_{\operatorname{GOE}}(x)dx=0$, we just need to consider $\varrho^{c}_{\operatorname{GOE}}(x,y)$, which is expressed as follows \cite{MethaBook}
\begin{equation}
	\varrho_{\mathrm{GOE}}^c(x, y)=-\frac{1}{2 \sigma^2}\left[s(r)^{2}-G(r)I(r)+\frac{1}{2}G(r)\right],
\end{equation}	
where $r=x-y$, $s(r)=\frac{\sin (\sqrt{N}\frac{r}{\sigma})}{r/\sqrt{2\sigma^{2}}}$, $I(r)=-\frac{\operatorname{Si}\left(\sqrt{N} \frac{r}{\sigma}\right)}{\pi}$, $G(r)=\frac{2 \sigma^2}{\pi} \frac{d}{d r}\left[\frac{\sin \left(\sqrt{N} \frac{r}{\sigma}\right)}{r}\right]$, and $\operatorname{Si}(t)$ is the sine integral function.
We set $w=x+y$, and change the integral variable $dxdy$ to $\frac{1}{2}dwdr$, obtaining
\begin{align}
		&\int xy\varrho_{\operatorname{GOE}}^{c}(x,y)dxdy\simeq -\frac{\sigma^{2}\pi^{2}N^{2}}{12}. \label{res0}
\end{align}
Finally, we have 
\begin{align}
		\langle D_L \rangle_{\operatorname{GOE}}&\simeq
		2\Gamma\sigma^{2}\left(N-2+\frac{\pi^{2}}{12}\right) \simeq 2 \Gamma \sigma^2 N.
\end{align}
Thus, the decoherence rate limit averaged over GOE is proportional to the Hilbert space dimension $N$, a feature termed extreme decoherence \cite{Xu19}. 

	The GUE and GSE have the same spectral density function as the GOE, given in Eq. (\ref{pdf}), with the only difference being the two-point correlation function \cite{MethaBook}. In the large dimension approximation, they are given by $\varrho_{\operatorname{GUE}}(x,y)=\varrho_{\operatorname{GUE}}(x)\varrho_{\operatorname{GUE}}(y)-\frac{1}{\sigma^{2}}\left[\frac{\sin (\sqrt{N}r\sigma^{-1})}{\pi r\sigma^{-1}}\right]^{2}$ and  $\varrho_{\operatorname{GSE}}(x,y)=\varrho_{\operatorname{GSE}}(x)\varrho_{\operatorname{GSE}}(y)-\frac{S^{2}(2\sqrt{2}r)}{2\sigma^{2}}+\frac{G(2\sqrt{2}r)[I(2\sqrt{2}r)-1/2]}{2\sigma^{2}}$, respectively. Employing analogous methods, we find that the averaged decoherence rate limits for both the GUE and GSE are identical
\begin{equation}
		\langle D_L \rangle_{\operatorname{GUE}}=\langle D_L \rangle_{\operatorname{GSE}}=\langle D_L \rangle_{\operatorname{GOE}}, \label{DtAll}
\end{equation}
as shown in Appendix \ref{app-A}. Crucially, the decoherence rate limit in all the GXE scales with the Hilbert space dimension $N$. This observation generalizes the finding for random dephasing dynamics in the GUE case \cite{Xu19}. The decoherence rate and its limit are thus independent of the underlying symmetry class and governed by the second moment of the ensemble from which Lindblad operators are sampled. In this sense, the decoherence rate of random Lindbladians exhibits universality. The GXEs are constructed to have a common second moment; thus, Eq. (\ref{DtAll}) holds. Deviations from this identity are to be expected in non-Gaussian ensembles in which other potentials govern the probability density for sampling a given matrix $L$. 
In addition, the scaling of the decoherence rate with the Hilbert space dimension $N$ holds, provided that the variance of the ensemble and $\Gamma$ are independent of $N$. Naturally, this dependence can be suppressed if these parameters are accordingly rescaled with $N$.

\begin{figure}[!t]
		\centering
		
		\subfigure{
			\begin{overpic}[width=0.99\linewidth]{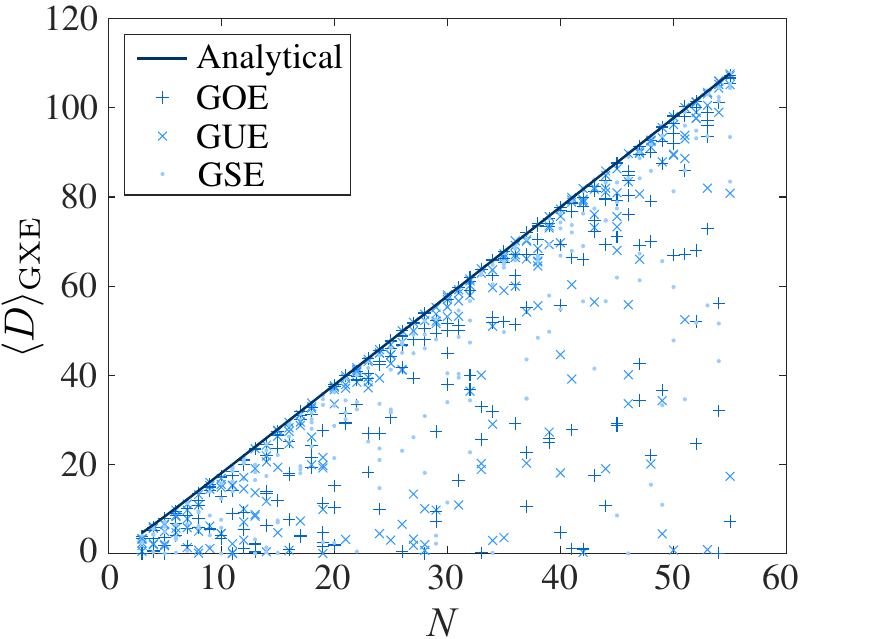}
			\end{overpic}
		}
\caption{Decoherence rate $\langle D\rangle_{\text {GXE }}$ as a function of the dimension of the Hilbert space $N\in[3,55]$, for three Gaussian ensembles: GOE, GUE, and GSE. For each dimension $N$, five initial states are chosen at random. Lindblad operators are sampled from GOE, GUE, and GSE, using 10,000 realizations with $\Gamma\sigma^2=1$. The analytical prediction (solid line) provides an upper bound for the numerical results, indicating a proportionality between the decoherence rate and the dimension of Hilbert space for random initial pure states.}
		\label{fig1}
\end{figure}

\begin{figure}[t]
	\centering
	\includegraphics[width=0.99\linewidth]{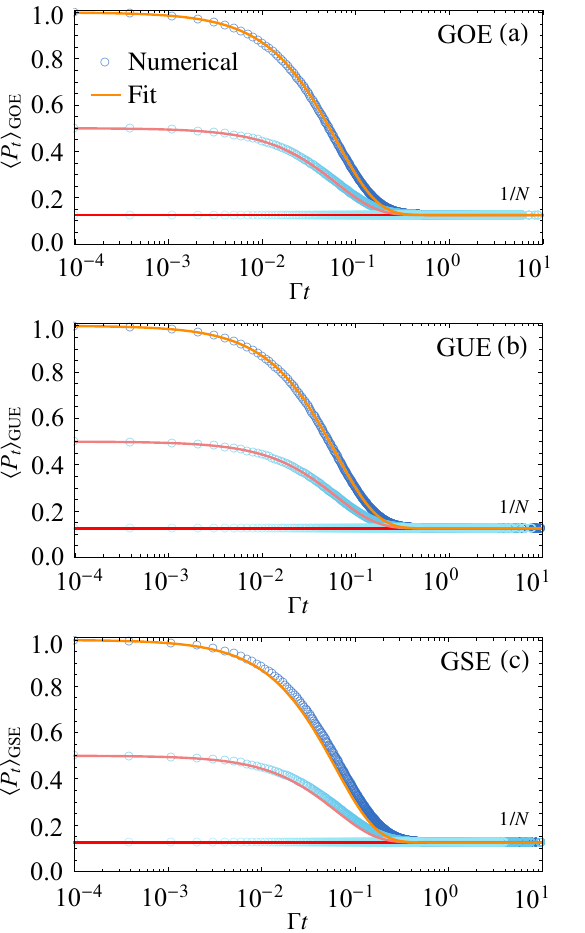}
	\caption{Logarithmic plot of purity as a function of time, employing random Lindblad operators averaged over GOE, GUE, and GSE, with $N=8$ and $\Gamma\sigma^2=1$. The study examines three initial state conditions: $P_0=1$, $1/2$, and $1/N$. Numerical evaluation: Each state undergoes analysis through 500 ensemble realizations. The solid curves correspond to the fitted function, i.e., Eq. (\ref{PGXEfit}), utilizing the decoherence rate limit $\langle D_L\rangle$.}
	\label{figfitGXE}	
\end{figure}

	To illustrate our findings, we present numerical simulations in Fig. \ref{fig1}. We randomly generated five initial states for each $N$ and ensemble type. The decoherence rate was subsequently averaged across GXEs, with $10,000$ realizations for each $N$. Our analysis demonstrates that the decoherence rate for pure initial states is proportional to the Hilbert space dimension $N$, thereby establishing a theoretical upper bound, i.e., a decoherence rate limit. The decoherence rate remains below this theoretical limit for all other initial states. We note that Hermitian Lindblad operators arise in a variety of contexts, e.g., from random phase changes on a short timescale \cite{Milburn1991}, stochastic fluctuations in the Hamiltonian \cite{Xu19,Chenu17}, the quantum evolution involving realistic clocks of finite precision \cite{Egusquiza1999,Egusquiza03},  randomized measurement schemes \cite{Korbicz17}, non-Hermitian Hamiltonian deformations \cite{Roubeas2023}, and spectral filtering in quantum chaotic systems \cite{Apollonas2023PRA}. As a result, it follows from Eq. (\ref{DtAll}) that the decoherence rate remains unaffected by the system's symmetries in these scenarios. 
	
Additionally,  while the decoherence rate is identified from the short-time behavior of purity, we next show that it effectively governs the decay of the purity averaged over ensembles at all times. This is in contrast with the expectation that other time scales arise in the dynamics \cite{Lidar98}.  As illustrated in Fig. \ref{figfitGXE}, $\left < P_t \right>_{\text{GXE}}$ exhibit a monotonic decay, characteristic of unital maps \cite{Lidar06}, eventually reaching a plateau with $P_\infty:=\left < P_{t\to \infty} \right>_{\text{GXE}}=1/N$, corresponding to the value of the purity in the maximal mixed state.  This motivates the following ansatz for the  purity as a function of time,  using the decoherence rate limit,
\begin{equation}
  P_{\text {fit}}=\left(P_0-P_\infty \right)\exp\left[-\langle D_L \rangle t \right]+P_\infty. \label{PGXEfit}
\end{equation}
Notably, in the GSE numerics depicted in Fig. \ref{figfitGXE}, the matrix dimension is maintained identical to that of GOE and GUE. The accuracy of the fit (\ref{PGXEfit}) in reproducing the numerically exact results for the purity decay at all times underscores the importance of the decoherence rate and its limit, on which we focus: it characterizes the dynamics of the purity in a simple way when the Lindblad operator is a random Hermitian matrix.

\section{The decoherence rate with GinXE} \label{SecGinXE}

\begin{figure}[t]
		\centering
		\subfigure{
		\begin{overpic}[width=0.99\linewidth]{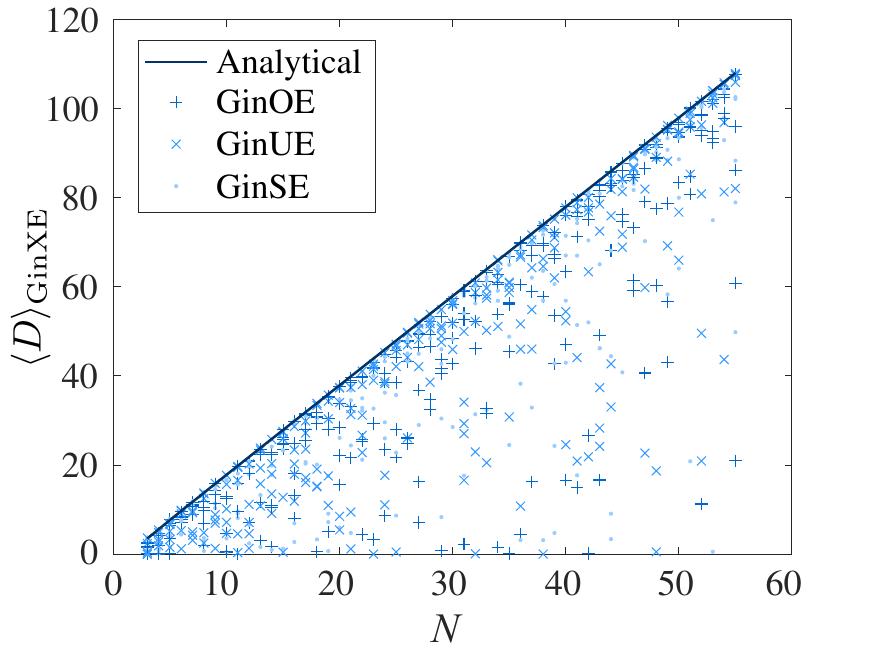}
		\end{overpic}
		}
		\caption{Decoherence rate, $\langle D\rangle_{\text {GinXE}}$, as a function of the Hilbert space dimension, $N\in [3,55]$, for three Ginibre ensembles: GinOE, GinUE, and GinSE. The analytical prediction for GinUE (solid line) suggests an upper bound for the decoherence rate, highlighting a direct proportionality with the Hilbert space dimension for random initial pure states.}
		\label{fig2}
\end{figure}

\begin{figure}[t]
		\centering
		\subfigure{
		\begin{overpic}[width=0.99\linewidth]{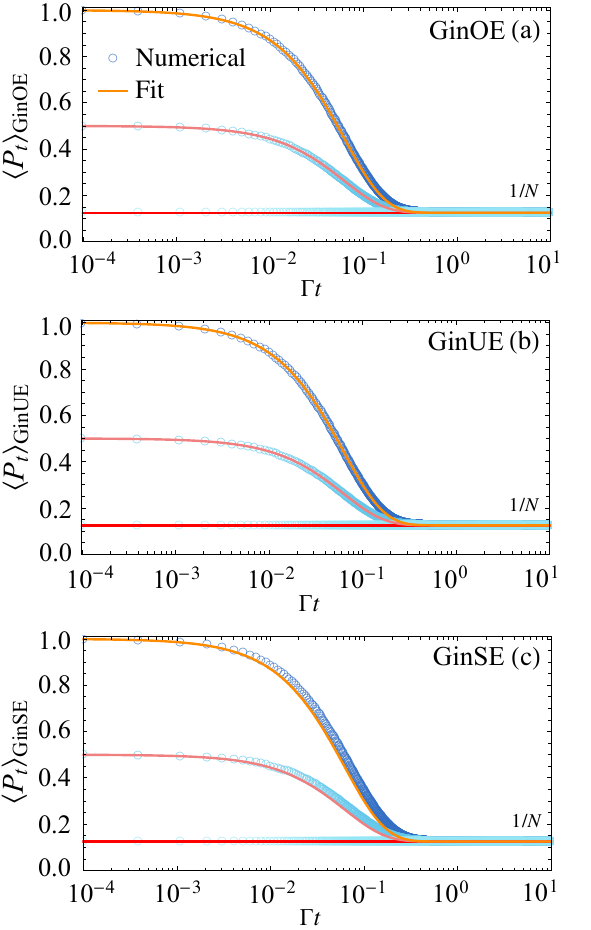}
		\end{overpic}
		}
		\caption{Logarithmic plot of purity as a function of time, employing random Lindblad operators averaged over GinOE, GinUE, and GinSE, with $N=8$ and $\Gamma\sigma^2=1$. Other parameters employed are consistent with those presented in Fig. \ref{figfitGXE}.}
	\label{figfitGinXE}	
\end{figure}

For random non-Hermitian Lindblad operators, we first consider a square $N\times N$ matrix $G$ with its matrix elements specified as $G_{mn} = e^{0}_{mn} + i e^{1}_{mn}$, where these elements are independent and identically distributed complex Gaussian random variables. This ensemble is commonly referred to as the complex Ginibre ensemble (GinUE) \cite{ginibre1965statistical}. 
%GinUE matrices display bi-unitary invariance, implying that they remain unchanged when subjected to the transformation $G\rightarrow UGV$, where both $U$ and $V$ represent arbitrary unitary matrices. 
Utilizing the Schur decomposition $G=U(\Lambda+T)U^{-1}$ \cite{horn1985matrix}, with $U$ as a unitary matrix, $\Lambda$ as a diagonal matrix containing its eigenvalues along the diagonal, and $T$ as a strictly upper triangular matrix, Eq. (\ref{upD}) reduces to 
%
%\begin{widetext}
%\begin{equation}
%\langle D_L \rangle_{\mathrm{GinUE}} = \frac{2 \Gamma}{N+1} \left[ \left\langle\operatorname{tr}\left(\Lambda^{\dagger} \Lambda + T^{\dagger} T\right)\right\rangle_{\mathrm{GinUE}} - \frac{\left\langle\operatorname{tr}\left(\Lambda^{\dagger}\right) \operatorname{tr}(\Lambda)\right\rangle_{\mathrm{GinUE}}}{N} \right]. \label{ginue}
%\end{equation}
%\end{widetext}
\begin{align}
\langle D_L \rangle_{\mathrm{GinUE}} & = \frac{2 \Gamma}{N+1} \bigg[  \left\langle\operatorname{tr}\left(\Lambda^{\dagger} \Lambda+T^{\dagger} T\right)\right\rangle_{\mathrm{GinUE}} \nonumber \\
& - \frac{\left\langle\operatorname{tr}\left(\Lambda^{\dagger}\right) \operatorname{tr}(\Lambda)\right\rangle_{\mathrm{GinUE}}}{N}\bigg]. \label{ginue}
\end{align}
According to the circular law, the spectral density function of GinUE follows a uniform distribution on a disk, i.e., $\varrho_{\operatorname{GinUE}}(z)=\frac{1}{\sigma^{2}\pi},|z|\leq\sigma\sqrt{N}$ \cite{MethaBook}. We can transform the integral variable $z$ from rectangular to polar coordinates in the complex plane for its analysis. Moreover, due to the bi-unitary invariance of GinUE, the distribution of $T$ remains unchanged, resulting in the same probability distribution function as the Gaussian complex random matrix $G$. Thus, the first term on the right-hand side of Eq. (\ref{ginue}) reads (see details in Appendix \ref{app-B}) 
	\begin{align}
		\left\langle\operatorname{tr}\left(\Lambda^{\dagger} \Lambda+T^{\dagger} T\right)\right\rangle_{\mathrm{GinUE}}=\frac{\sigma^2 N(2N-1)}{2}.
	\end{align}
	For addressing the second term in Eq. (\ref{ginue}), it is necessary to utilize the two-point correlation function $\varrho_{\operatorname{GinUE}}(z_{1},z_{2})=\pi^{-2}\sigma^{-4}\left[1-\operatorname{exp}(-|\frac{z_{1}-z_{2}}{\sigma}|^{2})\right]$ \cite{MethaBook}. To simplify the analysis, we initially transform the integral variables from $z_1$ and $z_2$ to $u=z_1-z_2$ and $v=z_1+z_2$, and subsequently to polar coordinates. This process ultimately yields 
	\begin{align}
		\langle\operatorname{tr}(\Lambda^{\dagger})\operatorname{tr}(\Lambda)\rangle_{\operatorname{GinUE}}
		\simeq&-\frac{N^{2}\sigma^{2}}{2}+2\sigma^{2}N.
	\end{align}
	Finally, the decoherence rate limit averaged over GinUEs is given by
\begin{equation}
\langle D_L \rangle_{\operatorname{GinUE}}
		\simeq 2\Gamma\sigma^{2}\frac{N^{2}-2}{N+1}\simeq 2\Gamma\sigma^{2}N.\label{res}
\end{equation}

Analytical calculations for GinOE and GinSE are cumbersome and pose significant challenges. Consequently, we turn to numerical simulations to conduct our analysis. In Fig.  \ref{fig2}, we depict the decoherence rate averaged over GinXEs as a function of 
$N$. Similar to the GXE cases, we average the decoherence rate over $10,000$ realizations for each $N$ and each GinXE type. As indicated by Eq. (\ref{res}), our analytical result establishes an upper bound for the decoherence rate averaged over GinXEs. In addition, this result indicates that the decoherence rate, proportionate to the  Hilbert space dimension, matches the same value obtained in the three GXEs. This indicates that the decoherence rate limit of random Lindbladians is independent of the type of Lindblad operators, regardless of whether they are Hermitian or non-Hermitian.

Similar to the analysis in Sec. \ref{SecGXE}, Fig. \ref{figfitGinXE} illustrates the time-dependent purity for GinXEs, exemplified with $N=8$. The solid curves depict the fitted function of purity, as defined by Eq. (\ref{PGXEfit}), which demonstrates a monotonically decreasing function of time before stabilizing at a plateau of at $P_{\infty}\to 1/N$. Accordingly, the dynamical behavior of purity influenced by non-Hermitian random Lindblad operators is also well described by an exponential decay governed by the decoherence rate limit.

\section{Concentration of decoherence rate}	\label{SecCP}

In the preceding sections, we observed a tendency for the decoherence rate to cluster near its upper limit; see Figs. \ref{fig1} and \ref{fig2}. This section is dedicated to a more detailed analysis of the underlying causes of this concentration.

Consider the system initially prepared in 
\begin{equation}
  \rho_0=\frac{1-p}{N}\mathbbm{1}+\ket{\Psi}\bra{\Psi} p,\label{rho0}
\end{equation}
where $\ket{\Psi}$ is an arbitrary pure state and $p=\sqrt{(N P_0-1)/(N-1)}$. Based on the results presented in Sec. \ref{SecGXE} and Sec. \ref{SecGinXE}, Eq. (\ref{<D>}) simplifies to the following form:
\begin{equation}
  \langle D\rangle=\Gamma \sigma^2 A \left( N-\frac{1}{P_0}  \right), \label{Dvariable}
\end{equation}	 
where $A=2+(\pi^2/6-2)/N$ for GXE and $A=2-2/(N^2-1)$ for GinXE. In the large $N$ limit, both entities converge towards the same value $A=2$. Let $P_0$ be a random variable uniformly distributed ranging from $1/N$ to $1$. Then the cumulative distribution function of the random variable $\langle D\rangle$, denoted as $F(\langle D\rangle)$, is given by 
\begin{equation}
  F(\langle D\rangle)=\frac{P_0-\frac{1}{N}}{1-\frac{1}{N}}=\frac{\langle D\rangle}{(N-1)(\tilde{A} N-\langle D\rangle)}, \label{CDF}
\end{equation}
where $\tilde{A}=\Gamma \sigma^2 A$ for short. Thus, the corresponding probability density function is expressed as 
\begin{equation}
  f(\langle D\rangle)=\frac{\partial F(\langle D\rangle)}{\partial \langle D\rangle}=\frac{\tilde{A} N}{N-1} \frac{1}{(\tilde{A} N-\langle D\rangle)^2}.\label{PDF}
\end{equation}

\begin{figure}[t]
		\centering
		\subfigure{
		\begin{overpic}[width=0.99\linewidth]{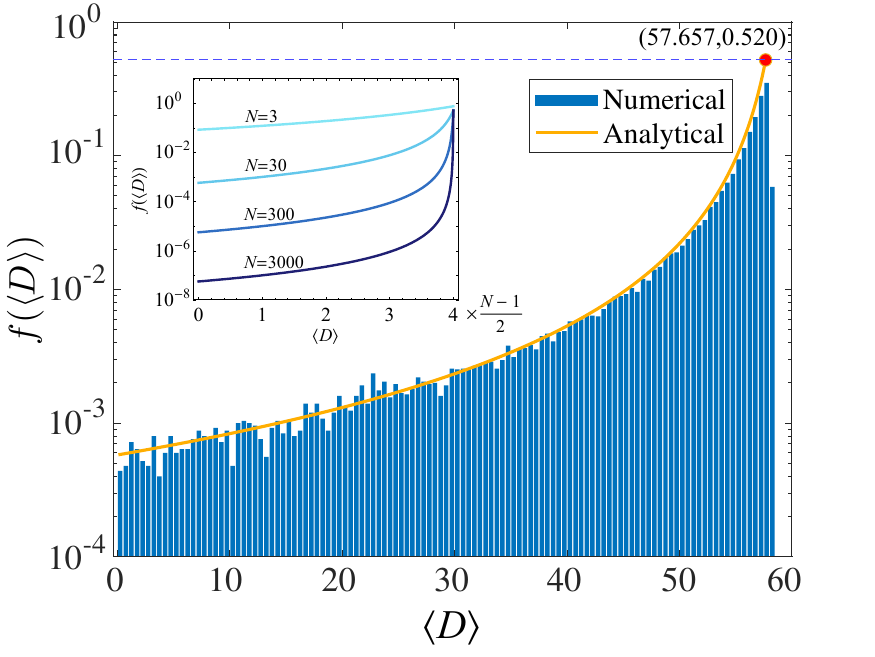}
		\end{overpic}
		}
		\caption{Logarithmic plot of the probability density function of the decoherence rate, denoted as $f(\langle D\rangle)$. This distribution was obtained through numerical simulations using the GOE as an example, with parameters $N=30$ and $\Gamma\sigma^2=1$. 50,000 initial states are prepared according to Eq. (\ref{Dvariable}) with $P_0$ uniformly distributed. For each initial state, the decoherence rate was averaged over 10,000 realizations of GOEs. The numerical results show good agreement with the analytical distribution, as described in Eq. (\ref{PDF}). The red dot represents the upper bound of the decoherence rate for $N=30$. Inset: $f(\langle D\rangle)$ versus $\langle D\rangle$ with $N$=$3$, $30$, $300$, and $3000$, respectively. A more pronounced concentration of decoherence rates is observed with the increasing Hilbert space dimension $N$.}
		\label{fig3}
\end{figure}

For illustration, Fig. \ref{fig3} depicts the probability density function, evaluated both analytically and numerically (using the GOE as an example). It has been rigorously established in former sections that the maximum decoherence rate is attained when the initial state is pure. Intriguingly, our findings here indicate a tendency for the decoherence rate to concentrate around its limit, a phenomenon that persists even when the initial states are mixed.

We note that our analytical results hold asymptotically in the limit of a large Hilbert space dimension. This explains why, in numerical simulations with a finite moderate value of the Hilbert space dimension, a small fraction of cases exceeds the predicted decoherence rate limit, as shown in Fig. \ref{fig3} when $N=30$.

\begin{figure}[t]
		\centering
		\subfigure{
		\begin{overpic}[width=1\linewidth]{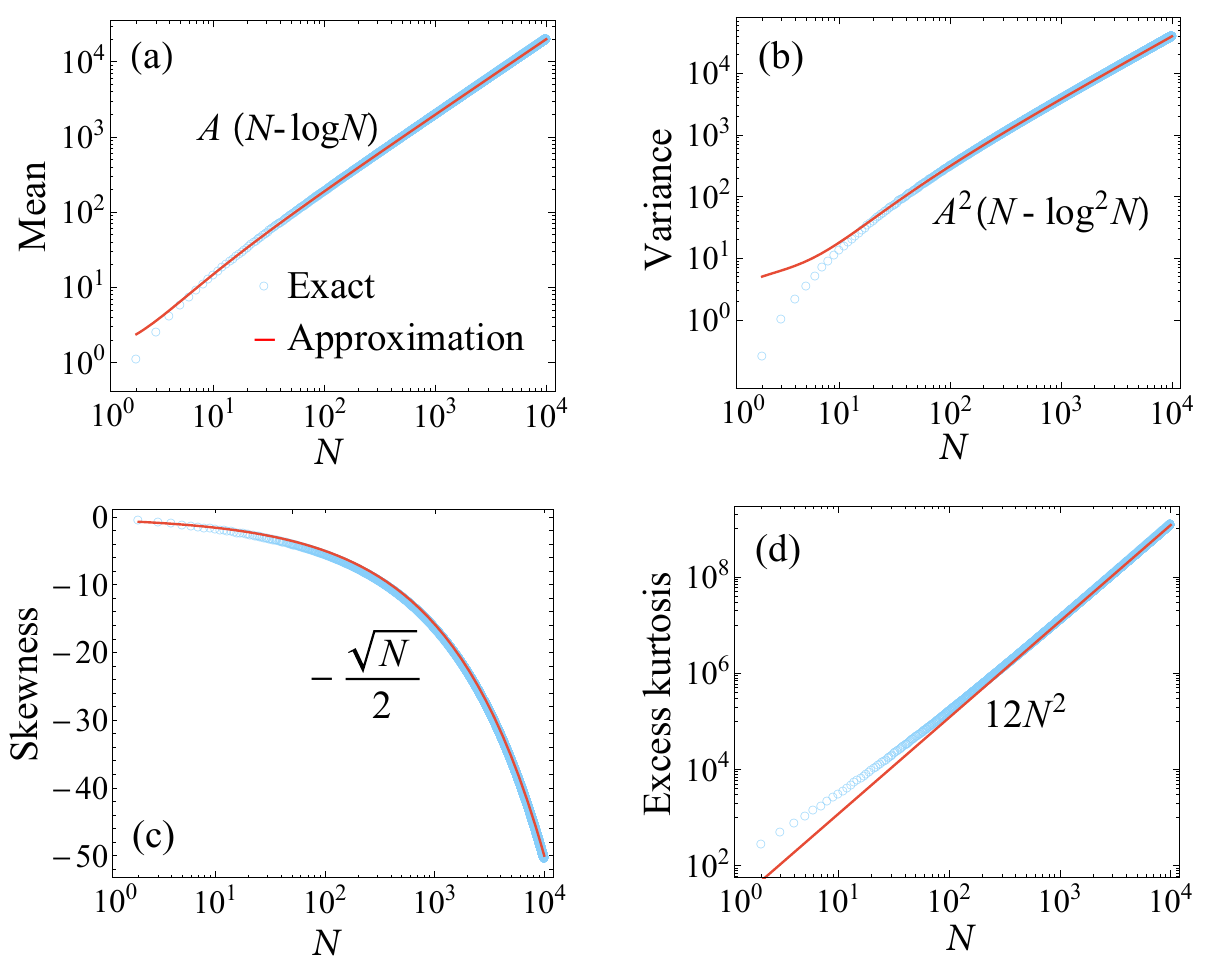}
		\end{overpic}
		}
		\caption{Depiction of mean, variance, skewness, and excess kurtosis as functions of the Hilbert space dimension, $N$, for the probability density function  $f(\langle D\rangle)$ of the decoherence rate with $\Gamma \sigma^2=1$. This graphical representation elucidates how these statistical parameters evolve with varying $N$, providing insights into the concentration of the decoherence rate. }
		\label{fig4}
\end{figure} 

The concentration of the ensemble-averaged decoherence rate can be more comprehensively understood by analyzing the moments and cumulants of the corresponding probability density function in Eq. (\ref{PDF}). The $k$-th moment of the distribution can be directly derived from the characteristic function
\begin{equation}
  f(\omega)=\int_0 ^{\tilde{A}(N-1)}  f(\langle D\rangle)e^{i\omega \langle D\rangle}d\langle D\rangle, 
\end{equation}
by differentiating $k$ times with respect to $\omega$ and then evaluate the result at $\omega=0$. 
%Similarly,  the $k$-th  cumulant can be obtained form the generating function $\log (f(\omega))$ respectively . 
After some algebra, the $k$-th moment function $E[\langle D\rangle^k]$ is given by

\begin{equation}
  E[\langle D\rangle^k] =\tilde{A}(N-1)^k \left[1-\frac{{}_2F_1 \left(1,1+k;2+k;\frac{N-1}{N}\right) k}{N(1+k)}  \right],\label{Mkth}
\end{equation}
where ${ }_2 F_1(a, b ; c ; z)=\sum_{n=0}^{\infty} \frac{(a)_n(b)_n}{(c)_n} \frac{z^n}{n !}$ is the hypergeometric function. Similarly, the $k$-th cumulant  $\kappa_k$ can be obtained using the expansion of the cumulant generating function $\log (f(\omega))=\sum_{k=1}^\infty \kappa_k (i\omega)^k/k!$ or directly constructed using Eq. (\ref{Mkth}). The first four cumulants are explicitly given by 
\begin{eqnarray}
%\begin{equation}
    \kappa_1&=&\tilde{A} N \left(1-\frac{\log N}{N-1}    \right), \nonumber\\%\label{k1}\\
%\end{equation}
%%
%\begin{equation}
    \kappa_2&=&\tilde{A}^2 N\left[1-\frac{N \log^2 N}{(N-1)^2}\right], \nonumber\\%\label{k2}\\
%\end{equation}
%
%\begin{equation}
    \kappa_3&=&-\frac{\tilde{A}^3 N}{2}\left[(N+1)-\frac{6N\log N}{N-1}+\frac{4N^2 \log^3 N}{(N-1)^3}\right], \nonumber\\
    %\nonumber\label{k3}\\
%\end{equation}
%and
%\begin{align}
\kappa_4 & =& \frac{\tilde{A}^4 N}{3} \bigg[ 1 + N(N - 8) - \frac{6N(N + 1)\log N}{N - 1} \nonumber \\
& &+ \frac{36N^2 \log^2 N}{(N - 1)^2} - \frac{18N^3 \log^4 N}{(N - 1)^4} \bigg]. \label{k4}
%\end{align}
\end{eqnarray}
Using them,  we are ready to characterize statistical measures such as the mean ($\kappa_1$), variance ($\kappa_2$), skewness ($\kappa_3/\kappa_2^{3/2}$), and excess kurtosis ($\kappa_4/\kappa_2^{2}$). In Fig. \ref{fig4}, we present these statistical measures as functions of the Hilbert space dimension $N$. These figures provide a multifaceted view of the concentration effect, most notably in the skewness and excess kurtosis, which exhibits a proportionality to the negative square root of $N$ and the square of $N$. Given that the mean is already proportional to $N$, such a pattern in the excess kurtosis suggests a tighter clustering of decoherence rates around the mean. These observations collectively indicate that with increasing $N$, the distribution increasingly concentrates at the upper bound of the decoherence rate, highlighting a key feature of the system's decoherence behavior with a large Hilbert space dimension.

\section{Discussion and Conclusion} \label{SecDC}
In the preceding sections, we have derived the limits to the decoherence rates across typical random matrix ensembles with Wigner-Dyson symmetry classes, all of which share a common upper bound and the concentration phenomenon of the decoherence rate. This enables us to explore the decoherence rate for Lindblad operators described by mixing random matrices originating from identical or distinct ensembles, such as the combination of GXE and GYE, or the direct product of GXE and GYE, where $\operatorname{X}, \operatorname{Y}\in\{\operatorname{O},\operatorname{U},\operatorname{S}\}$. Such an analysis may hold significant implications for the study of the low-energy limit of symmetry breaking in quantum chromodynamics with GXE-GYE crossover \cite{Kanazawa2020,KANAZAWAPLB}.
	For instance, consider a random Lindblad operator with the following forms: $ L_{\alpha} = a_1 L_{\alpha}^{1} + a_2 L_{\alpha}^{2}$, where $a_{1,2}$ are normalization constants, and $L_{\alpha}^{1}$ and $L_{\alpha}^{2}$ are arbitrary matrices belonging to the Gaussian ensembles. The decoherence rate is given by $D=\sum_{k,l=1}^{2} D_{kl}$, where $D_{kl}=(2a_k a_l/P_0) \sum_{\alpha}\gamma_{\alpha}\left[ \operatorname{tr}(\rho_{0}^2 L_{\alpha}^{k\dagger}L_{\alpha}^{l})-\operatorname{tr}(\rho_{0}L_{\alpha}^{k\dagger}\rho_{0}L_{\alpha}^{l})\right]$.
%	\begin{equation}
%		D=\sum_{k,l=1}^{2} D_{kl},
%	\end{equation}
%where
%	\begin{equation}
%		D_{kl}=\frac{2a_k a_l \sum_{\alpha}\gamma_{\alpha}\left[ \operatorname{tr}(\rho_{0}^2 L_{\alpha}^{k\dagger}L_{\alpha}^{l})-\operatorname{tr}(\rho_{0}L_{\alpha}^{k\dagger}\rho_{0}L_{\alpha}^{l})\right]}{P_0}.
%	\end{equation}
For $D_{12}$ and $D_{21}$, given the absence of correlations between two distinct types of random matrices, it follows that $\langle D_{12}\rangle=\langle D_{21}\rangle=0$. Furthermore, $D_{11}$ and $D_{22}$ are exactly equal to the decoherence rates of Gaussian ensembles $\langle D_{11}\rangle=a_{1}^{2}\langle D\rangle_{\operatorname{GXE}}$ and $\langle D_{22}\rangle=a_{2}^{2}\langle D\rangle_{\operatorname{GYE}}$. Since $\langle D\rangle_{\operatorname{GXE}}=\langle D\rangle_{\operatorname{GYE}}$ and $a_{1}^{2}+a_{2}^{2}=1$, the decoherence rate limit of $L_{\alpha}$ is also proportional to the dimension of the Hilbert space.

To conclude, we have established the universality of the decoherence rate limit for open chaotic quantum systems, employing Lindblad operators drawn from both typical Hermitian and non-Hermitian random matrices. The observed uniformity in decoherence rates, which scale linearly with the Hilbert space dimension, generalize the scenario of extreme decoherence reported in the GUE scenario \cite{Xu19}. Therefore, the fastest decoherence rate under Markovian dynamics is reached by random Lindblad dynamics, regardless of the specific symmetry considered. 
Furthermore, in an example of a one-parameter family of initial quantum states, our research illustrates that those open quantum chaotic systems, characterized by random Lindbladians, invariably tend to exhibit, in a statistical sense, the fastest decoherence, clustering near the limit, for almost any initial state.
Finally, our findings demonstrate the stability of the decoherence rate even when combining matrices from distinct ensembles. 

These findings are of relevance to quantum foundations and the understanding of decoherence in complex quantum systems in relation to dissipative quantum chaos. Likewise, they can be applied to various aspects of quantum technologies, quantum information, and conformal field theory in the presence of environmental errors \cite{Xu19,delCampo2020,GG2022PRD}.

\section*{Acknowledgements}
The authors are indebted to Lucas S\'a, Federico Balducci, and Pablo Azcona for insightful comments on the manuscript. This work was supported by the National Natural Science Foundation of China under Grant No.12074280 and by the Luxembourg National Research Fund (FNR), grant reference 17132054. For open access, the authors have applied a Creative Commons Attribution 4.0 International (CC BY 4.0) license to any Author Accepted Manuscript version arising from this submission.
	
%\newpage

\setcounter{secnumdepth}{2} %Numbering in the Appendix
\appendix\label{app}
\onecolumngrid 

\section{Ensemble Average over Gaussian Random Matrices} \label{app-0}
For simplicity, we temporarily omit the subscript $\alpha$ from $L_{\alpha}$ in the subsequent discussion and initially focus on the unitary ensembles. The calculation of the averaged decoherence rate, with the Lindblad operator $L\in$ GUE, involves the quantities  $\langle L^{\dagger}L\rangle$ and $\langle L^\dagger \rho_{0}L\rangle$. These can be expressed in terms of the Haar average of the function of $L$ \cite{Xu19}, i.e.,
\begin{equation}
  \langle f(L)\rangle=\int \langle f(L)\rangle_{\text {Haar}} \mathcal{D}x, \label{Average}
\end{equation}
where $\mathcal{D} x=\varrho\left(x_1, \ldots, x_N\right) \prod_l d x_l$ denotes as the integration measure, $\varrho\left(x_1, \ldots, x_N\right)$ is the $N$-point joint probability density function of the random matrix, and
\begin{equation}
  \langle f(L)\rangle_{\text {Haar}}=\int f\left(U L U^{-1}\right) d \mu(U).\label{Haar}
\end{equation}
Here, $d \mu(U)$ is the uniform probability (Haar) measure on the unitary group. The integrations over the Haar measure in the expressions for $\langle L^{\dagger}L\rangle_{\text {Haar}}$ and $\langle L^\dagger \rho_{0}L\rangle_{\text {Haar}}$ in Eq. (\ref{Haar}) are given by 
\begin{equation}
    \langle L^\dagger L\rangle_{\text {Haar}}=\int U L^\dagger L U^{-1} d \mu(U)=\frac{\operatorname{tr}\left(L^{\dagger} L\right)}{N} \mathbbm{1},\label{LL}
\end{equation}
and
\begin{equation}
     \langle L^\dagger \rho_0 L\rangle_{\text {Haar}}=\int U L^\dagger U^{-1}\rho_{0} U L U^{-1} d\mu(U)=\frac{1}{N(N^2-1)}\left[ \operatorname{tr}\left(L^{\dagger} L\right) \left(N \mathbbm 1- \rho_0 \right)- \operatorname{tr}\left(L^{\dagger}\right) \operatorname{tr}\left(L\right) \left(\mathbbm 1-N \rho_0 \right)   \right], \label{LrL}
\end{equation}
where we have utilized the second and fourth-moment functions of the unitary group \cite{collins2006integration,Collins_2009,akemann2011oxford,gessner2013generic}:
\begin{equation}
    \int U A U^{-1} d \mu(U)=\frac{\operatorname{tr}(A)}{N} \mathbbm{1},\label{second}
\end{equation}
and 
\begin{equation}
    \int U A U^{-1} X U B U^{-1} d \mu(U)=\frac{1}{N(N^2-1)}\left[ \operatorname{tr}\left(AB\right) \left(N \mathbbm 1- X \right)- \operatorname{tr}\left(A\right) \operatorname{tr}\left(B\right) \left(\mathbbm 1-N X \right)   \right]. \label{forth}
\end{equation}
Substituting Eqs. (\ref{LL}) and (\ref{LrL}) into Eq. (\ref{Average}), we have 
\begin{equation}
    \langle L^\dagger L\rangle= \frac{\left < \operatorname{tr}(L^\dagger L) \right>}{N} \mathbbm{1}, \label{LLaveraged}
\end{equation}
and
\begin{equation}
    \langle L^\dagger \rho_0 L\rangle=\frac{1}{N(N^2-1)}\left[ \left <\operatorname{tr}\left(L^{\dagger} L\right) \right > \left(N \mathbbm 1- \rho_0 \right)- \left <\operatorname{tr}\left(L^{\dagger}\right) \operatorname{tr}\left(L\right) \right >\left(\mathbbm 1-N \rho_0 \right)   \right]. \label{LrLaveraged}
\end{equation}
Then, the GUE averaged decoherence rate in Eq. (\ref{D}) reads
\begin{equation}
    \langle D\rangle=2 \Gamma \frac{N P_0-1}{\left(N^2-1\right) P_0}\left[\left\langle\operatorname{tr}\left(L^{\dagger} L\right)\right\rangle-\frac{1}{N}\left\langle\left(\operatorname{tr} L^{\dagger}\right)(\operatorname{tr} L)\right\rangle\right]. \label{DV}
\end{equation}

For $L$ belonging to GinUE, the Schur decomposition $L=U(\Lambda+T)U^{-1}$ can be utilized \cite{horn1985matrix}. Here, $U$ is a unitary matrix, $\Lambda$ is a diagonal matrix with eigenvalues along its diagonal, and $T$ is a strictly upper triangular matrix. Equations (\ref{Average})-(\ref{DV}) remain valid; the only modification required is the replacement of the integration measure with $\mathcal{D}x=\varrho (x_{1},\ldots,x_{N})\left[\prod_{m<n}g(x_{mn})dx_{mn}\right]\prod_{l}dx_{l}$. Here $g(\cdot)$ is the function with Gaussian distribution. The above analysis is also extendable to other Gaussian and Ginibre ensembles with Wigner-Dyson symmetry classes, based on the integration with respect to the Haar measure on compact Lie groups, like orthogonal and symplectic groups \cite{collins2006integration}. Given that our calculations are only based on the second and fourth moments of the corresponding group, identical to those in Eqs. (\ref{second}) and (\ref{forth}), the final outcome, i.e., Eq. (\ref{DV}), remains applicable. This will be further substantiated by exact numerical analysis in Secs. \ref{SecGXE} and \ref{SecGinXE}.

\section{Regarding the Traceless of Lindblad Operators} \label{app-C}
For simplicity, the main text employs the Lindblad operator $L$ directly sampled from GXE or GinXE, which deviates from the definition of being traceless. Here, we prove that the results remain equivalent despite this deviation. To facilitate this demonstration, we begin with the general Eq. (\ref{<D>}) and define the traceless Lindblad operator $\tilde{L}$ as

\begin{equation}
 \tilde{L}=L-\frac{\mathbbm{1}}{N}\operatorname{tr}L, \label{TracelessL}
\end{equation}
with which we have 

\begin{equation}
\left\langle\operatorname{tr}\left(L^{\dagger} L\right)\right\rangle=\left\langle\operatorname{tr}\left(\tilde{L}^\dagger+\frac{\mathbbm{1}}{N}\operatorname{tr}L^\dagger \right)\left( \tilde L+\frac{\mathbbm{1}}{N}\operatorname{tr}L\right)\right\rangle=\left\langle\operatorname{tr}\left(\tilde L^{\dagger} \tilde L\right)\right\rangle+\frac{1}{N}\left\langle\operatorname{tr}\left(L^{\dagger}\right\rangle \operatorname{tr}\left( L\right)\right\rangle.\label{TL1}
\end{equation}
Thus,
\begin{equation}
\left\langle\operatorname{tr}\left(L^{\dagger} L\right)\right\rangle-\frac{1}{N}\left\langle\left(\operatorname{tr} L^{\dagger}\right)(\operatorname{tr} L)\right\rangle=\left\langle\operatorname{tr}\left(\tilde{L}^{\dagger} \tilde{L}\right)\right\rangle,
\end{equation}
which completes our proof.

\section{Decoherence Rates Averaged Over GXEs} \label{app-A}
Previous studies have employed varying standard deviations, leading to differences in the probability density function ($\mathit{pdf}$) and two-point correlation function \cite{MethaBook,BookVivo2017,Cotler2017,Xu2021}. Let the $\mathit{pdf}$ and two-point correlation function in the case with standard deviation 1 be denoted by $\varrho^{(1)}(x)$ and $\varrho^{(1)}(x,y)$,  in contrast with   $\varrho^{(\sigma)}(x)$ and $\varrho^{(\sigma)}(x,y)$, for the case with standard deviation $\sigma$. Because of $\int\varrho^{(1)}(x)dx=\int\varrho^{(\sigma)}(x)dx$ and $\int\varrho^{(1)}(x,y)dxdy=\int\varrho^{(\sigma)}(x,y)dxdy$, we have the formula $\varrho^{(\sigma)}(x)=\frac{1}{\sigma}\varrho^{(1)}(\frac{x}{\sigma})$ and $\varrho^{(\sigma)}(x,y)=\frac{1}{\sigma^{2}}\varrho^{(1)}(\frac{x}{\sigma},\frac{y}{\sigma})$. Thus, the first term and the second term on the right-hand side of Eq. (\ref{upD}) averaged over GOE, i.e., $\langle D\rangle_{\operatorname{GOE}}$, are
		\begin{equation}
			\langle \operatorname{tr}(L^{2})\rangle_{\operatorname{GOE}}=\int_{-2\sigma\sqrt{N}}^{2\sigma\sqrt{N}} x^{2}\varrho_{\operatorname{GOE}}(x)dx
			=\sigma^{2}N^{2},
		\end{equation}
and
		\begin{equation}
			\langle\operatorname{tr}(L)^{2}\rangle_{\operatorname{GOE}}=\int x^{2}\varrho_{\operatorname{GOE}}(x)dx+\int xy\varrho_{\operatorname{GOE}}(x,y)dxdy.
		\end{equation}
The two-point correlation function of GOE has the following form \cite{MethaBook}
		\begin{equation}
			\varrho_{\operatorname{GOE}}(x,y)=\frac{1}{2\sigma^{2}}\operatorname{det}  \begin{bmatrix}
				K_{N}(x,x) & K_{N}(x,y) \\
				K_{N}(y,x) & K_{N}(y,y)
			\end{bmatrix} ,
		\end{equation}
where $K_{N}(x,y)$ is a quaternion, its matrix form is
\begin{equation}
	K_{N}(x,y)=\begin{bmatrix}
		S_{N}(x,y) & G_{N}(x,y) \\
		I_{N}(x,y)-\frac{1}{2}\operatorname{sign}(x-y) & S_{N}(y,x)
	\end{bmatrix}.
\end{equation}
Here the sign function $\operatorname{sign}(t)$ takes values  1, $-1$ and 0 when $t>0$, $t<0$ and $t=0$, respectively.
 We divide the two-point correlation function into two parts $\varrho_{\operatorname{GOE}}(x,y)=\varrho_{\operatorname{GOE}}(x)\varrho_{\operatorname{GOE}}(y)+\varrho^{c}_{\operatorname{GOE}}(x,y)$. The first term here is equal to zero after integration and
		\begin{equation}
			\varrho^{c}_{\operatorname{GOE}}(x,y)=-\frac{1}{2\sigma^{2}}S_{N}(x,y)S_{N}(y,x)+\frac{1}{2\sigma^{2}}\left[I_{N}(x,y)-\frac{1}{2}\operatorname{sign}(x-y)\right]G_{N}(x,y).
		\end{equation}
The approximations of $S_{N}(x,y)$ in the limit of large dimension are
		\begin{equation}
			\lim_{N\rightarrow\infty}S_{N}(x,y) =\frac{\sin (\sqrt{N}\frac{r}{\sigma})}{r/\sqrt{2\sigma^{2}}}= s(r),
		\end{equation}
		\begin{equation}
			\lim_{N\rightarrow\infty}\operatorname{sign}(x-y)G_{N}(x,y) =\frac{2\sigma^{2}}{\pi}\frac{d}{dr}\left[ \frac{\sin (\sqrt{N}\frac{r}{\sigma})}{r}\right]
			= G(r),
		\end{equation}
		\begin{equation}
			\lim_{N\rightarrow\infty}\operatorname{sign}(x-y)I_{N}(x,y) =-\frac{\operatorname{Si}(\sqrt{N}\frac{r}{\sigma})}{\pi}
			= I(r),
		\end{equation}
where $r=x-y$ and $\operatorname{Si}(t)$ is sine integral function. Thus,
		\begin{equation}
			\lim_{N\rightarrow\infty}\varrho_{\operatorname{GOE}}^{c}(x,y)=-\frac{1}{2\sigma^{2}}\left[s(r)^{2}-G(r)I(r)+\frac{1}{2}G(r)\right].
		\end{equation}
Changing the integral variable, we get
\begin{equation}
\int x y \varrho_{\mathrm{GOE}}^c(x, y) d x d y=-\int_{-\pi \sigma \sqrt{N}}^{\pi \sigma \sqrt{N}} \int_{-\pi \sigma \sqrt{N}+|r|}^{\pi \sigma \sqrt{N}-|r|} -\frac{w^2-r^2}{8}\left[s(r)^{2}-G(r)I(r)+\frac{1}{2}G(r)\right] d r d w \simeq -\frac{\sigma^{2}\pi^{2}N^{2}}{12},
\end{equation}
where we used $\lim_{t\rightarrow\infty}\operatorname{Si}(t)=\frac{\pi}{2}$. Except for the first term, we can ignore other terms with powers less than two. Finally,
\begin{equation}
    \langle D_L\rangle_{\mathrm{GOE}} \simeq \Gamma \left(\frac{2 \sigma^2 N^2}{N+1}-\frac{2 \sigma^2 N^2}{N(N+1)}+\frac{\pi^2 \sigma^2 N^2}{6 N(N+1)}\right)\simeq 2 \Gamma \sigma^2\left(N-2+\frac{\pi^2}{12}\right)\propto 2 \Gamma \sigma^2 N.
\end{equation}

For the GOE and GSE,
\begin{equation}
    \varrho_{\mathrm{GUE}}^c(x, y)=-\frac{1}{\sigma^2}\left[\frac{\sin \left(\sqrt{N} r \sigma^{-1}\right)}{\pi r \sigma^{-1}}\right]^2,
\end{equation}
and
\begin{equation}
    \varrho_{\mathrm{GSE}}^c(x, y)=-\frac{s^2(2 \sqrt {2}r)}{2 \sigma^2}+\frac{G(2 \sqrt{2}r)[I(2\sqrt{2} r)-1 / 2]}{2 \sigma^2}.
\end{equation}
As a result, 
		\begin{equation}
			\int xy\varrho_{\operatorname{GUE}}(x,y)dxdy
			\simeq-\frac{\sigma^{2}}{48}\left[8\pi\operatorname{Si}(2\pi N)N^{2}+4N-4\pi N\right.
			+\left.4\cos (2\pi N)N+O(N^{2})\right] ,
		\end{equation}
		\begin{equation}
			\lim_{N\rightarrow\infty}\frac{1}{N(N+1)}\langle \operatorname{tr}(L)^{2}\rangle_{\operatorname{GUE}}=\left(1-\frac{\pi^{2}}{12}\right)\sigma^{2},
		\end{equation}
		\begin{equation}
			\int xy\varrho_{\operatorname{GSE}}(x,y)dxdy 
			\simeq - \frac{\sigma^{2}\pi^{2}N^{2}}{12},
		\end{equation}
and
		\begin{equation}
			\lim_{N\rightarrow\infty}\frac{1}{N(N+1)}\langle \operatorname{tr}(L)^{2}\rangle_{\operatorname{GSE}}=\left(1-\frac{\pi^{2}}{12}\right)\sigma^{2}.
		\end{equation}
Finally, we obtain the decoherence rate limits of the GUE and GSE
		\begin{equation}
			\langle D_L \rangle_{\operatorname{GUE}}=\langle D_L \rangle_{\operatorname{GSE}}\simeq 2\Gamma\sigma^{2}\left(N-2+\frac{\pi^{2}}{12}\right). \label{res1}
		\end{equation}
In summary, the decoherence rates for both the GUE and the GSE align with those of the GOE, with each being proportional to the dimension of the system's Hilbert space.

\section{Decoherence Rates Averaged Over GinUEs} \label{app-B}	
The decoherence rate limit for GinUE is calculated as follows. Using the circular law $\varrho_{\operatorname{GinUE}}(z)=\frac{1}{\sigma^{2}\pi}, |z|\leq\sigma\sqrt{N}$ \cite{MethaBook} and the bi-unitary invariance of GinUE, the first term on the right-hand side of Eq. (\ref{ginue}) reads
\begin{equation}
			\left\langle \operatorname{tr}\left( \Lambda^{\dagger} \Lambda \right)  \right\rangle_{\text {GinUE }}
			=\int_{0}^{2\pi}d\theta\int_{0}^{\sigma\sqrt{N}}R^{3}\frac{1}{\sigma^{2}\pi}dR
			=\frac{\sigma^{2}N^{2}}{2},
\end{equation}
and
\begin{align}
\left\langle \operatorname{tr}\left( T^{\dagger} T \right)  \right\rangle_{\text {GinUE }} & =\sum_{i<j} \int \frac{\exp \left(-\left|\frac{T_{i j}}{\sigma}\right|^2\right)}{\sigma^2 \pi}\left|T_{i j}\right|^2 \frac{1}{2} d T_{i j}^* d T_{i j} \nonumber \\
& =\frac{N(N-1)}{2} \int \frac{\exp \left(-\left|\frac{\tau+i \delta}{\sigma}\right|^2\right)}{\sigma^2 \pi}\left|\tau+i \delta\right|^2 d \tau d \delta \nonumber \\
& =\frac{N(N-1)}{2} \times \int_0^{2 \pi} \int_0^{\infty} R^3 \frac{1}{\sigma^2 \pi} e^{-\frac{R^2}{\sigma^2}} d R d \theta \nonumber \\
& =\frac{\sigma^2 N(N-1)}{2},
\end{align}
where $T_{ij}=\tau+i \delta$. Next, we calculate the second term on the right-hand side of Eq. (\ref{ginue}): 
		\begin{equation}
			\langle\operatorname{tr}(\Lambda^{\dagger})\operatorname{tr}(\Lambda)\rangle_{\operatorname{GinUE}}
			=\int z_{1}z_{2}^{\ast}\rho_{\operatorname{GinUE}}(z_{1},z_{2})\frac{1}{4}dz_{1}dz_{1}^{\ast}dz_{2}dz_{2}^{\ast}
			+\int zz^{\ast}\rho_{\operatorname{GinUE}}(z)dzdz^{\ast},\label{B6}
		\end{equation}
where, under the large-dimension approximation, the two-point correlation function is given by $\rho_{\operatorname{GinUE}}(z_{1},z_{2})=\pi^{-2}\sigma^{-4}\left[1-\operatorname{exp}(-|\frac{z_{1}-z_{2}}{\sigma}|^{2})\right]$ \cite{MethaBook}. 
Changing the integration variables $u=z_{1}-z_{2}$ and $v=z_{1}+z_{2}$  yields
		\begin{equation}
			\int dz_{1}dz_{1}^{\ast}dz_{2}dz_{2}^{\ast}=\frac{1}{2}\int dudu^{\ast}dvdv^{\ast}.
		\end{equation}
Thus, 
		\begin{align}
			\int z_{1}z_{2}^{\ast}\rho_{\operatorname{GinUE}}(z_{1},z_{2})\frac{1}{4}dz_{1}dz_{1}^{\ast}dz_{2}dz_{2}^{\ast}
			=&\int \frac{z_{1}z_{2}^{\ast}}{4\pi^{2}\sigma^{4}}\left[1-\operatorname{exp}\left(-\left|\frac{z_{1}-z_{2}}{\sigma}\right|^{2}\right)\right]dz_{1}dz_{1}^{\ast}dz_{2}dz_{2}^{\ast}\nonumber \\
			=&\int \frac{|v|^{2}-|u|^{2}}{32\pi^{2}\sigma^{4}}\left[1-\operatorname{exp}\left(-\left|\frac{u}{\sigma}\right|^{2}\right)\right]dudu^{\ast}dvdv^{\ast}, \label{B8}
		\end{align}
where we have employed the integral symmetry		
		\begin{align}
			&\int z_{1}z_{2}^{\ast}\rho_{\operatorname{GinUE}}(z_{1},z_{2})dz_{1}dz_{1}^{\ast}dz_{2}dz_{2}^{\ast}\nonumber \\
			=&\int z_{1}^{\ast}z_{2}\rho_{\operatorname{GinUE}}(z_{1},z_{2})dz_{1}dz_{1}^{\ast}dz_{2}dz_{2}^{\ast}\nonumber \\
			=&\int z_{1}z_{2}\rho_{\operatorname{GinUE}}(z_{1},z_{2})dz_{1}dz_{1}^{\ast}dz_{2}dz_{2}^{\ast}\nonumber \\
			=&\int z_{1}^{\ast}z_{2}^{\ast}\rho_{\operatorname{GinUE}}(z_{1},z_{2})dz_{1}dz_{1}^{\ast}dz_{2}dz_{2}^{\ast},
		\end{align}
and $z_{1}z_{2}^{\ast}=(|v|^{2}-|u|^{2})/4$. In polar coordinates, Eq. (\ref{B8}) simplifies to
		\begin{align}
			\int z_{1}z_{2}^{\ast}\rho_{\operatorname{GinUE}}(z_{1},z_{2})\frac{1}{4}dz_{1}dz_{1}^{\ast}dz_{2}dz_{2}^{\ast} 
			&=\int_{0}^{2\pi}d\theta_{1}\int_{0}^{2\pi}d\theta_{2}\int_{0}^{2\sigma\sqrt{N}}\int_{0}^{\sqrt{4\sigma^{2}N-r_{2}^{2}}}\frac{r_{1}^{2}-r_{2}^{2}}{8\sigma^{4}\pi^{2}}r_{1}r_{2}(1-e^{(-\frac{r_{2}}{\sigma})^{2}})dr_{1}dr_{2}\nonumber \\
			&=\frac{\sigma^{2}e^{-4N}}{8}\left[3+4N+e^{4N}(-8N^{2}+8N-3)\right]\nonumber \\
			&\simeq-\sigma^{2}N^{2}+2\sigma^{2}N.
		\end{align}
Since $\int z z^* \rho_{\mathrm{GinUE}}(z) d z d z^*=N^2 \sigma^2/2$, Eq. (\ref{B6}) is given by
		\begin{equation}
			\langle\operatorname{tr}(\Lambda^{\dagger})\operatorname{tr}(\Lambda)\rangle_{\operatorname{GinUE}}
			\simeq-\sigma^{2}N^{2}+\frac{N^{2}\sigma^{2}}{2}+2\sigma^{2}N
			=-\frac{N^{2}\sigma^{2}}{2}+2\sigma^{2}N.
		\end{equation}
Finally, we obtain the decoherence rate limit of GinUE in the main text (see Eq. (\ref{res}))
		\begin{equation}
		\langle D_L \rangle_{\mathrm{GinUE}} \simeq 2 \Gamma\left[\frac{\sigma^2 N^2}{2(N+1)}+\frac{\sigma^2 N(N-1)}{2(N+1)}+\frac{\sigma^2 N-4 \sigma^2}{2(N+1)}\right] \lesssim 2 \Gamma \sigma^2 \frac{N^2-2}{N+1}\propto 2\Gamma\sigma^{2}N.
		\end{equation}

\twocolumngrid 
	
\bibliography{refsRM}

\end{document}